\input{psfig}
\documentstyle[onecolumn,doublespacing]{mn}

\newcommand{\be}{\begin{equation}}
\newcommand{\ee}{\end{equation}}
\newcommand{\bea}{\begin{eqnarray}}
\newcommand{\eea}{\end{eqnarray}}
\newcommand{\bc}{\begin{center}}

\newcommand{\ec}{\end{center}}
\def\spose#1{\hbox to 0pt{#1\hss}}
\newcommand{\lta}{\mathrel{\spose{\lower 3pt\hbox{$\mathchar"218$}}
     \raise 2.0pt\hbox{$\mathchar"13C$}}}
\newcommand{\gta}{\mathrel{\spose{\lower 3pt\hbox{$\mathchar"218$}}
     \raise 2.0pt\hbox{$\mathchar"13E$}}}

\def\H0{$H_0= 100~h~$km\,s$^{-1}$\,Mpc$^{-1}$}

\newif\ifAMStwofonts

\ifoldfss
  \ifCUPmtlplainloaded \else
    \NewTextAlphabet{textbfit} {cmbxti10} {}
    \NewTextAlphabet{textbfss} {cmssbx10} {}
    \NewMathAlphabet{mathbfit} {cmbxti10} {} 
    \NewMathAlphabet{mathbfss} {cmssbx10} {} 
  \fi
  \ifAMStwofonts
    \ifCUPmtlplainloaded \else
      \NewSymbolFont{upmath} {eurm10}
      \NewSymbolFont{AMSa} {msam10}
      \NewMathSymbol{\upi}     {0}{upmath}{19}
      \NewMathSymbol{\umu}     {0}{upmath}{16}
      \NewMathSymbol{\upartial}{0}{upmath}{40}
      \NewMathSymbol{\leqslant}{3}{AMSa}{36}
      \NewMathSymbol{\geqslant}{3}{AMSa}{3E}

      \let\leq=\leqslant 
       
    \fi
  \fi
\fi 

\ifnfssone
  \newmathalphabet{\mathit}
  \addtoversion{normal}{\mathit}{cmr}{m}{it}
  \addtoversion{bold}{\mathit}{cmr}{bx}{it}
  \newmathalphabet{\mathbfit} 
  \addtoversion{normal}{\mathbfit}{cmr}{bx}{it}
  \addtoversion{bold}{\mathbfit}{cmr}{bx}{it}
  \newmathalphabet{\mathbfss} 
  \addtoversion{normal}{\mathbfss}{cmss}{bx}{n}
  \addtoversion{bold}{\mathbfss}{cmss}{bx}{n}
  \ifAMStwofonts
    \ifCUPmtlplainloaded \else
and
your
      \UseAMStwoboldmath
      \makeatletter
      \new@mathgroup\upmath@group
      \define@mathgroup\mv@normal\upmath@group{eur}{m}{n}
      \define@mathgroup\mv@bold\upmath@group{eur}{b}{n}
      \edef\UPM{\hexnumber\upmath@group}
      \new@mathgroup\amsa@group
      \define@mathgroup\mv@normal\amsa@group{msa}{m}{n}
      \define@mathgroup\mv@bold\amsa@group{msa}{m}{n}
      \edef\AMSa{\hexnumber\amsa@group}
      \makeatother
      \mathchardef\upi="0\UPM19
      \mathchardef\umu="0\UPM16
      \mathchardef\upartial="0\UPM40
      \mathchardef\leqslant="3\AMSa36
      \mathchardef\geqslant="3\AMSa3E

      \let\leq=\leqslant 

    \fi
  \fi
\fi 

\ifnfsstwo
  \DeclareMathAlphabet{\mathbfit}{OT1}{cmr}{bx}{it}
  \SetMathAlphabet\mathbfit{bold}{OT1}{cmr}{bx}{it}
  \DeclareMathAlphabet{\mathbfss}{OT1}{cmss}{bx}{n}
  \SetMathAlphabet\mathbfss{bold}{OT1}{cmss}{bx}{n}

  \ifAMStwofonts
    \ifCUPmtlplainloaded \else
      \DeclareSymbolFont{UPM}{U}{eur}{m}{n}
      \SetSymbolFont{UPM}{bold}{U}{eur}{b}{n}
      \DeclareSymbolFont{AMSa}{U}{msa}{m}{n}
      \DeclareMathSymbol{\upi}{0}{UPM}{"19}
      \DeclareMathSymbol{\umu}{0}{UPM}{"16}
      \DeclareMathSymbol{\upartial}{0}{UPM}{"40}
      \DeclareMathSymbol{\leqslant}{3}{AMSa}{"36}
      \DeclareMathSymbol{\geqslant}{3}{AMSa}{"3E}

      \let\leq=\leqslant 

    \fi
  \fi
\fi 

\ifCUPmtlplainloaded \else
  \ifAMStwofonts \else 
    \def\upi{\pi}
    \def\umu{\mu}
    \def\upartial{\partial}
  \fi
\fi

\title{A theoretical study of the mass temperature relation for clusters of galaxies}
\author[A.Del Popolo et al.]
  {A. Del Popolo,$^1$$^,$$^2$$^,$$^3$\\
  $^1$ Dipartimento di Matematica, Universit\`{a} Statale di Bergamo,
  Piazza Rosate, 2 - I 24129 Bergamo, ITALY \\
     $^2$ Feza G\"ursey Institute, P.O. Box 6 \c Cengelk\"oy, Istanbul,
     Turkey\\
     $3$  Bo$\breve{g}azi$\c{c}i University, Physics Department,
     80815 Bebek, Istanbul, Turkey
}
\date{Accepted ???
      Received 2000 July 24;
      in original form ???}

\pagerange{\pageref{firstpage}--\pageref{lastpage}} \pubyear{2000}

\begin{document}

\maketitle

\label{firstpage}

\begin{abstract}

I derive the mass-temperature relation and its time evolution for clusters of galaxies 
in different cosmologies by means of two different models. The first one is a modification and improvement of a model by Del Popolo \& Gambera (1999),
namely based upon a modification of the top-hat model in order to take account of angular momentum acquisition by protostructures
and of an external pressure term in the virial theorem. 
The second one is based on the merging-halo formalism of Lacey \& Cole (1993), accounting for the fact that
massive clusters accrete matter quasi-continuously, and is an improvement of a model proposed by 
Voit (2000) (herafter V2000), again to take account of angular momentum acquisition by protostructures. 
The final result is that, in both models, the M-T relation shows a break at $T \sim 3-4 {\rm keV}$.
The behavior of the M-T relation is as usual, $M \propto T^{3/2}$, at the high mass end, and $M \propto T^{\gamma}$, with a value of $\gamma>3/2$ depending on the chosen cosmology. Larger values of $\gamma$ are related to open cosmologies, while $\Lambda$CDM cosmologies give results of the slope intermediate between the flat case and the open case. 
The evolution of the M-T relation, for a given $M_{\rm vir}$, is more modest both in flat and open universes in comparison to previous estimate found in literature, even more modest than what found by V2000.
Moreover the time evolution is more rapid in models with $L=0$ than in models in which 
the angular momentum acquisition by protostructures is taken into account ($L \neq 0$). The effect of a non-zero cosmological constant is that of slightly increasing the evolution of the M-T relation with respect to open models with $L \neq 0$. The evolution is more rapid for larger values (in absolute value) of the spectral index, $n$.\\
The mass-temperature relation, obtained using the quoted models, is also compared with the 
data by Finoguenov, Reiprich \& Bohringer (2001) (hereafter FRB). 
The comparison shows that the FRB data is 
able to rule out very low $\Omega_0$ models ($<0.3$), particularly in the 
open case, and that better fit are obtained   
by 
$\Lambda$CDM models and by CDM models with $\Omega_0>0.3$.
\end{abstract}



\section{Introduction}

In the past decade, observations of clusters of galaxies (e.g, ROSAT, ASCA) have shown the existence of a tight
correlation between the total gravitating mass of clusters, $M_{\rm tot}$, their X-ray luminosity ($L_{\rm X}$),
and temperature ($T_{\rm X}$) of the intra-cluster medium (ICM) (David et al. 1993; Markevitch 1998; Horner, 
Mushotzky \& Sharf 1999 (hereafter HMS)). 
The importance of these relations is due to the fact that cluster masses are difficult to measure directly, and when comparing cluster observations with models of structure formation a surrogate for cluster mass is usually used. Since $M_{\rm tot}$ compares with the ICM temperature measurements which can be obtained through X-ray spectroscopy, this explains the importance of an M-T relation.
Numerical simulations and theoretical studies have motivated the existence of such kind 
of relations, 
whose comparison can 
provide strong constraints on the prevailing cosmological models. 

For what concerns N-body simulations, in particular, Evrard, Metzler \& Navarro (1996), and Eke, 
Navarro \& Frenk (1997) showed that there is a tight relation between the mass of a cluster and its global X-ray temperature
in gas-dynamical simulations, irrespective of the assumed cosmological model and of the state of the cluster. 
Other N-body simulations by Mathiesen \& Evrard (2001) have shown that the ICM X-ray temperatures show a tight ($\leq 20 \%$ scatter) correlation with the total cluster mass components, correlation also seen  
in observations (Mohr, Mathiesen \& Evrard 1999).
 
From the theoretical point of view, within the framework of pure gravitational infall (e.g., Lilje 1992) by means of a top-hat model and simple arguments, based on the 
virialization density, suggest a self-similar relation, $M_{\rm tot} \propto T_{\rm X}^{\alpha}$ \footnote{$\alpha=3/2$ is almost unrelated to the settings of the cosmological parameters} confirmed by the results of N-body simulations (Evrard, Metzler \& Navarro 1996; Eke et al. 1998). The proportionality coefficient of the law has been usually fixed by using numerical simulations to calibrate it (Evrard et al. 1996; Thomas et al. 2001). Afshordi \& Cen (2001) (hereafter AC), through of energy conservation and assuming nearly spherical collapse, were able to obtain a normalization factor in agreement with hydro-simulations, but about 50\% lower than X-ray mass estimates (see also V2000) (this discrepancy with observational normalizations is probably due to some effect, happening at small scales (AC), which the model does not take into account).
For sake of completeness, it is important to stress that the results from different observational methods of mass measurement are not consistent with one another and with simulation results (e.g., HMS; Nevalainen, Markevitch \& Forman 2000 (hereafter NMF), FRB). While mass estimates from galaxy velocity dispersion are consistent with simulations (HMS), X-ray mass estimates 
can differ by 80\%
from the prediction of hydro-simulations. 

Another drawback of the standard top-hat model has been stressed by Voit \& Donahue (1998) (hereafter V98) and V2000. Using the merging-halo formalism of Lacey \& Cole (1993), which accounts for the fact that massive clusters accrete matter quasi-continuously, they showed that the M-T relation evolves, with time, more modestly than what expected in previous models 
and this evolution is even more modest in open universes. 

Finally, there is another problem that, added to the previously quoted ones, 
urges for a better theoretical understanding of the M-T relation: 
recent studies have shown that the self-similarity in the M-T relation seems to break at 
some keV (NMF; Xu, Jin \& Wu 2001). By means of ASCA data, using a small sample of 9 clusters (6 at 4 keV and 3 at $\sim 1$ keV), NMF has shown that $M_{\rm tot} \propto T_{\rm X}^{1.79 \pm 0.14}$ for the whole sample, and 
$M_{\rm tot} \propto T_{\rm X}^{3/2}$ excluding the low-temperature clusters. Xu, Jin \& Wu (2001) has found $M_{\rm tot} \propto T_{\rm X}^{1.60 \pm 0.04}$ (using the $\beta$ model), and $M_{\rm tot} \propto T_{\rm X}^{1.81 \pm 0.14}$ by means of the Navarro, Frenk \& White (1995) profile. FRB have investigated the T-M relation in the low-mass end finding that $M \propto T^{\sim 2}$, and $M \propto T^{\sim 3/2}$ at the high mass end. This behavior has been attributed to the effect of the formation redshift (FRB) (but see Mathiesen 2001 for a different point of view), or to cooling processes (Muanwong et al. 2001) and heating (Bialek, Evrard \& Mohr 2000). AC has shown that non-sphericity introduces an asymmetric, mass dependent, scatter for the M-T relation altering its slope at the low mass end ($T \sim 3$ keV).\\
In this paper, I derive the mass-temperature relation, and its time evolution, for clusters of galaxies, 
in different cosmologies by means of two different models: the first one is a modification and improvement of a model by Del Popolo \& Gambera (1999), based upon a modification of the top-hat model in order to take account of angular momentum acquisition by protostructures and of an external pressure term in the virial theorem.
The second one is 
an improvement of a model proposed by V2000, again to take account of angular momentum acquisition by protostructures. 
Both models show that the M-T relation is not self-similar. A  break is present in the quoted relation at $T \sim 3 {\rm keV}$ and at the lower mass end the power law index of the M-T relation is larger than $\alpha=3/2$ even in flat universes. 
The slope of the power-law index depends on the considered cosmology.
The two models also agree in predicting a more modest time evolution of the quoted relation in comparison with the results of previous models, and again depending on the cosmology.
%

The plan of the paper is the following: in section~(2) I describe the top-hat modified model (hereafter THM) and in section~(3) the modified V2000 model (hereafter VM). Section~(4) describes the results and section~(5) is devoted to conclusions.

\section{Review of the Top-Hat model}

As previously quoted, numerical methods and simple scaling arguments suggest that the X-ray temperature of clusters, $T_{\rm X}$, can be directly related to their masses. In fact, assuming that the mean density within a radius $r$ is $\Delta$ times the critical density, $\rho_{\rm b}$, one can express the mass in the given radius as: $M(\Delta) \propto T_{\rm X}^{3/2} \rho_{\rm b}^{-1/2} \Delta^{-1/2}$. Defining the quantity $\Delta_{\rm vir}$ as the density contrast of a spherical top-hat perturbation just after collapse and virialization, in the ``vein" of the Press-Schechter approach, the mass can be expressed as $M_{\rm vir}=M(\Delta_{\rm vir})$ and so $M_{\rm vir} \propto T_{\rm X}^{3/2} \rho_{\rm b}^{-1/2} \Delta_{\rm vir}^{-1/2}$, which gives the quoted relation between mass and temperature.  

In the following, I'll use Del Popolo \& Gambera (1999)
in order to get the M-T relation. 
The model is fundamentally a modification of the top-hat model in order to take account of angular momentum acquisition by protostructures and uses a modified version of the virial theorem in order to include a surface pressure term (V2000, AC). This correction is due to the fact that at the virial radius $r_{\rm vir}$ the density is non-zero and this requires a surface pressure term to be included in the virial theorem (Carlberg, Yee \& Ellingson 1997) (the existence of this confining pressure is usually not accounted for in the top-hat collapse model). 

The equation governing the collapse of a density perturbation taking account 
angular momentum acquisition by protostructures 
can be obtained using a model due to Peebles (Peebles 1993) (see also Del Popolo \& Gambera 1998, 1999).\\
 Let's consider an ensemble of gravitationally growing mass concentrations 
and suppose that the material in each system collects within the
same potential well
with inward pointing acceleration given by $g(r)$ (see Del Popolo \& Gambera 1998). We
indicate with $dP=f(L,r v_r,t)dL dv_r dr$ the probability that a particle
can be found  in the proper radius range $r$, $r+dr$, in the radial
velocity range $v_r={\dot r}$, $v_r+d v_r$ and with angular momentum
$L=r v_\theta$ in the range $dL$.
The radial
acceleration of the particle is: 
\begin{equation}
\frac{dv_r}{dt}=\frac{L^2(r)}{M^2r^3}-g(r)=\frac{L^2(r)}{M^{2} r^3}-\frac{G M }{r^2}  
\label{eq:col}
\end{equation}
Eq. (\ref{eq:col}) can be derived from a potential
and then from Liouville's theorem it follows that
the distribution function, $f$,
satisfies the collisionless Boltzmann equation:
\begin{equation}
\frac{\partial f}{\partial t} + v_{r}
\frac{\partial f}{\partial r} + \frac{\partial f}{\partial v_{r}}
\cdot \left[ \frac{L_{2}}{r^{3}} - g(r) \right] = 0
\end{equation}

Assuming a non-zero cosmological constant Eq. (\ref{eq:col}) 
becomes:
\begin{equation}
\frac{dv_r}{dt}=-\frac{G M }{r^2}+\frac{L^2(r)}{M^{2} r^3}+\frac{\Lambda}{3} r \label{eq:coll}
\end{equation}
(Peebles 1993; Bartlett \& Silk 1993; Lahav 1991; Del Popolo \& Gambera 1998, 1999).
Integrating Eq. (\ref{eq:coll}) we have: 
\begin{equation}
\frac{1}{2}\left( \frac{dr}{dt}\right) ^{2}=\frac{GM}{r}+\int 
\frac{L^{2}}{M^{2}r^{3}}dr+\frac{\Lambda }{6}r^{2}+\epsilon
\label{eq:coll1}
\end{equation}
where the value of the specific binding energy of the shell, $\epsilon$, can be obtained using the condition for turn-around, $\frac{dr}{dt}=0$.

The CDM spectrum used in this paper is that of Bardeen et al. (1986) 
normalized 
to reproduce the observed abundance of rich cluster of galaxies (e.g., Bahcal \& Fan 1998) 
Filtering the quoted spectrum on clusters scales, $R_f=3h^{-1}Mpc$, I obtained the
total angular momentum
acquired during expansion,
as described in Del Popolo \& Gambera (1998, 1999) (more hints on the model and some of the model limits can be found in 
Del Popolo, Ercan \& Gambera (2001)). 

The temperature-mass relation can be obtained using the virial theorem, energy
conservation and using Eq. (\ref{eq:coll1}). In Eq. (\ref{eq:coll1}), three forms of potential are present and the virial theorem can be written in the form:
\begin{equation}
\langle K \rangle=-\frac{1}{2} \langle U_{\rm G} \rangle-\langle U_{\rm L} \rangle+\langle U_{\Lambda}\rangle
\label{eq:virial}
\end{equation}
(Landau \& Lifshitz 1960; Lahav et al. 1991, Del Popolo \& gambera 1999), 
where, using Bartlett \& Silk (1993) notation, 
the meaning of $\langle \rangle$ is the time-average of the quantity considered,
$U_{\rm G}$ the gravitational potential energy, $U_{\rm L}$ that connected to the angular momentum and $U_{\Lambda }$ that due to $\Lambda$. 
\begin{figure}
\label{Fig. 1} \centerline{\hbox{
Fig. (1)
\psfig{file=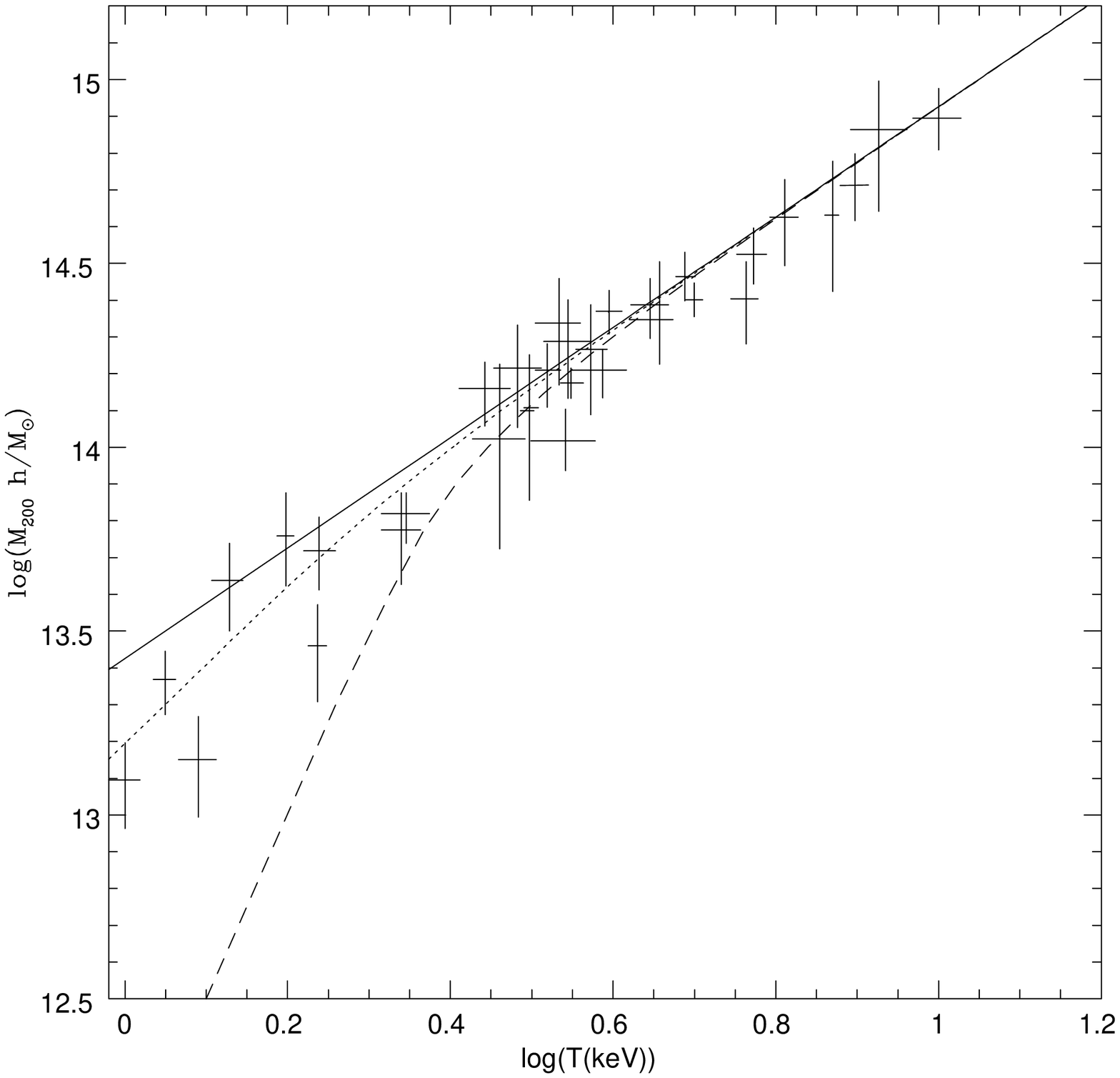,width=9cm} Fig. (2)
\psfig{file=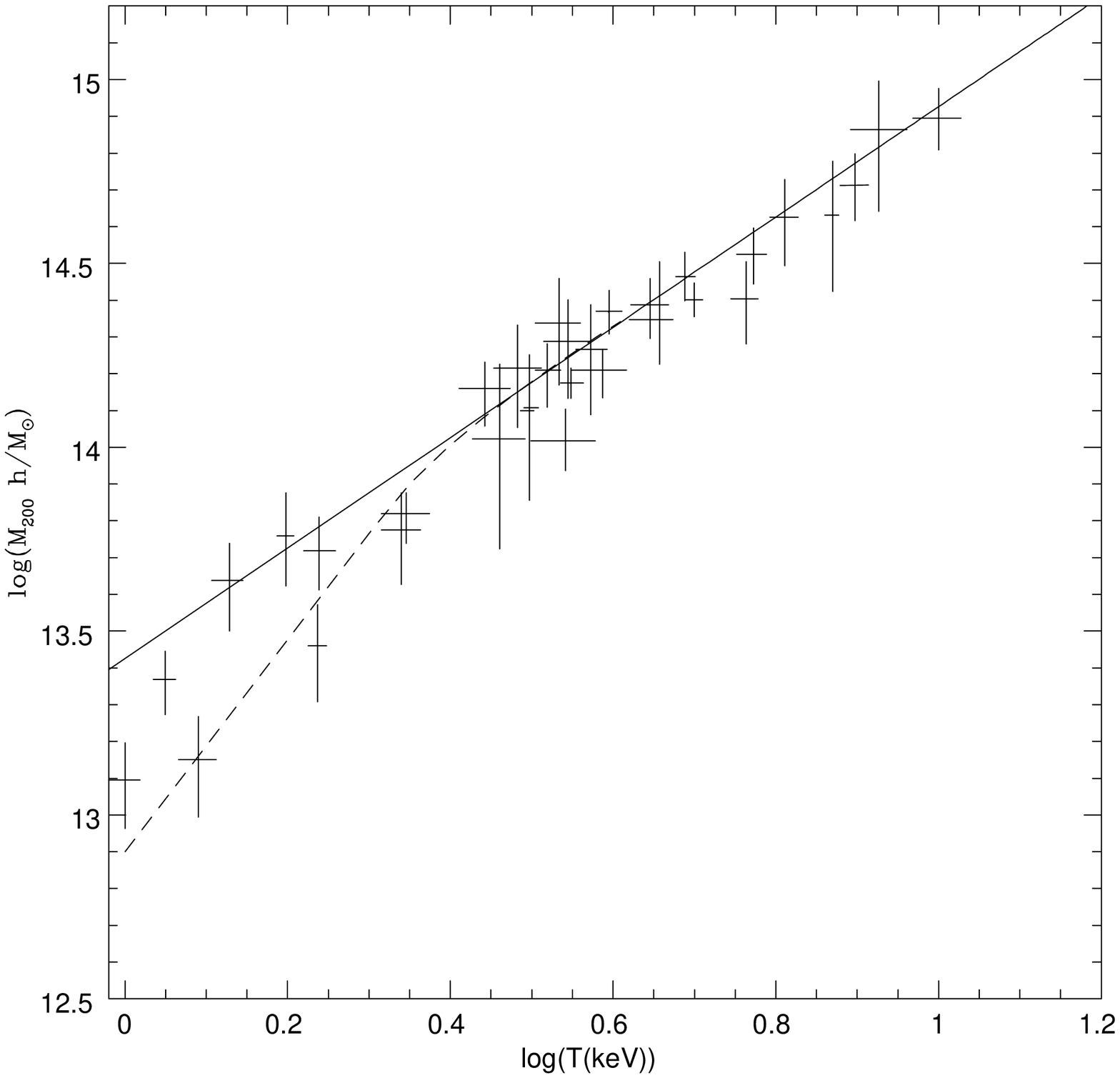,width=9cm}
}}
\caption[]{M-T relation predicted by the modified top-hat model. 
The solid line is the prediction of Eq. (\ref{eq:temp}), for $\Omega_{\Lambda}=0$, $\Omega_{\rm 0}=1$, shifted downwards, similarly to AC, to fit the FRB observational data in the massive end. 
The dotted and dashed line represents the prediction of the same equation
for $L \neq 0$, $\Omega_{\Lambda}=0$, $\Omega_{\rm 0}=1$ and $L \neq 0$, $\Omega_{\Lambda}=0$ and $\Omega_{\rm 0}=0.3$, respectively. 

{\bf Figure 2.}
As Fig. 1 but now $\Omega_{\Lambda} \neq 0$. The solid line is equivalent to the solid one in Fig. 1. The dashed line represents the case of a flat universe with $L \neq 0$, $\Lambda \neq 0$, ($\Omega_{\rm 0}=0.3$, $\Omega_{\Lambda}=0.7$).
}
\end{figure}
%
%
For example, in the case of the gravitational potential energy, we can write $U_{\rm G}=-\frac{G M}{r_{\rm eff}}$, where $r_{eff}$ is the time-averaged radius of a mass shell. 
As previously mentioned, when a surface pressure term is added, Eq. (\ref{eq:virial}) must be modified. Assuming that: 
\begin{equation}
\langle K \rangle+\langle E \rangle=3 P_{\rm ext} V 
\end{equation}
(AC),
where $P_{\rm ext}$ is the pressure of the outer boundary of virialized region and V is the volume, and by using the same assumption of AC, namely:
\begin{equation}
3 P_{\rm ext} V= -\nu U 
\end{equation}
where $\nu$ is a coefficient and $U$ is the total potential. With the introduction of the surface term Eq. (\ref{eq:virial}) becomes:
\begin{equation}
\langle K \rangle=-\frac{\nu+ 1}{2} \langle U_{\rm G}\rangle-(\nu+1) \langle U_{\rm L} \rangle+(\nu+1) \langle U_{\Lambda } \rangle
\label{eq:virial1}
\end{equation}

Energy conservation requires that (see Lahav 1991, Del Popolo \& Gambera 1999):
\begin{equation}
\langle E \rangle= \langle K \rangle+ \langle U_{\rm G} \rangle+\langle U_{\rm \Lambda} \rangle+\langle U_{\rm L} \rangle=U_{\rm G,ta}+U_{\rm \Lambda,ta}+U_{\rm L,ta}
\label{eq:energy}
\end{equation}
where the subscript ``${\rm ta}$" stands for turn-around. 
The previous equation by means of Eq. (\ref{eq:virial1}) reads:
\begin{equation}
\frac{1-\nu}{2}\langle U_{\rm G} \rangle+(2+\nu) \langle U_{\rm \Lambda} \rangle-\nu \langle U_{\rm L} \rangle=U_{\rm G,ta}+U_{\rm \Lambda,ta}+U_{\rm L,ta}
\label{eq:virial2}
\end{equation}
Eq. (\ref{eq:energy}) and Eq. (\ref{eq:virial1}) or Eq. (\ref{eq:virial2}), allows us to solve for $K$ and $r_{\rm eff}$. 

In fact Eq. (\ref{eq:virial1}) can be written as:
\begin{equation}
\langle K \rangle=-\left( \frac{\nu +1}{2}\right) U_{\rm G}\left[ 1
+2\frac{U_{\rm L}}{U_{\rm G}}-2\frac{U_{\rm \Lambda }}{U_{\rm G}}
\right] =\left( \frac{\nu +1}{2}\right) \frac{GM}{r_{\rm eff}}\left[ 1+2\frac{r_{\rm eff}}{GM^{3}}\int_{0}^{r_{\rm eff}}\frac{L^{2}}{r^{3}}dr-\frac{\Lambda r_{\rm eff}^{3}}{3GM}\right] 
\label{eq:virial20}
\end{equation}
where $r_{eff}= \psi r_{ta}$, $r_{ta}$ being the radius of the turn-around epoch. 
Using the expressions: 
\begin{equation}
M=4 \pi \rho_{\rm b} x^{3}_{1}/3
\label{eq:massb}
\end{equation}
$ \xi=r_{ta}/x_{1}$, $\Omega_0=\frac{8 \pi G \rho_{\rm b}}{3 H_0^2}$, I get:
\begin{equation}
r_{eff}=\psi \xi \left( \frac{2GM}{\Omega _{0}H_{0}^{2}}\right)^{1/3}
\end{equation}
Inserting this last equation in Eq. (\ref{eq:virial20}), I obtain:
\begin{equation}
\langle K \rangle=
\frac{(\nu+1)}{2}\frac{(G M  H_0)^{2/3} \Omega_0^{1/3}}{2^{1/3} \psi \xi}
\left[ 1+2^{\frac{4}{3}}\frac{\psi \xi }{G^{\frac{2}{3}}\Omega _{0}^{\frac{1}{3}}H_{0}^{\frac{2}{3}}M^{\frac{8}{3}}}\int_{0}^{r_{eff}}\frac{L^{2}}{r^{3}}dr-\frac{2\Lambda \left( \psi \xi \right) ^{3}}{3\Omega _{0}H_{0}^{2}}\right] 
\label{eq:virial21}
\end{equation}
Substituting $G=\frac{3H_{0}^{2}\Omega _{0}}{8\pi \rho _{b}}$ in Eq. (\ref{eq:virial21}), I finally get:
%
%
%
\begin{equation}
\langle K \rangle = 
%
\frac{(\nu+1)}{2}\frac{(G M  H_0)^{2/3} \Omega_0^{1/3}}{2^{1/3} \psi \xi}\left[
1+\frac{2^{\frac{10}{3}}\pi ^{\frac{2}{3}}\psi \xi 
\rho _{b}^{\frac{2}{3}}}{3^{\frac{2}{3}}
\Omega_0 H_0^{2}M^{\frac{8}{3}}}\int_0^{r_{\rm eff}} \frac{L^{2}}{r^{3}}dr-\frac{2}{3}
\Lambda \frac{\left( \psi \xi \right) ^{3}}{\Omega_0 H_0^{2}}
\right]
\label{eq:vifi}
\end{equation}
%
%
where $\Omega_{\Lambda}=\frac{\Lambda}{3 H_0^2}=1-\Omega_0$. 
Recalling the definitions of $\psi$, $\xi$, Eq. (\ref{eq:massb}), and  
since at tun-around we have $M=4 \pi \rho_{\rm ta} r_{\rm ta }^3/3$, it is possible to express the 
product $\psi \xi$ as:
\begin{equation}
\psi \xi=\frac{r_{\rm eff}}{r_{\rm ta}} \frac{r_{\rm ta}}{x_1}=\frac{r_{\rm eff}}{r_{\rm ta}}
\left(\frac{\rho_{\rm b}}{\rho_{\rm ta}}\right)^{1/3} (1+z_{\rm ta})^{-1} 
\label{eq:psixi}
\end{equation}
(see also Eq. (\ref{eq:zdep}) and Lilje (1992)).

In order to get a connection between the kinetic energy and temperature, I utilize the usual relation:
\begin{equation}
\langle K \rangle= \frac{3 \tilde{\beta} M k T}{2 \mu m_{\rm p}}
\label {eq:conn} 
\end{equation}
(AC),
where $k$ is the Boltzmann constant, $\mu=0.59$ is the mean molecular weight, $m_{\rm p}$ the proton mass and $\tilde{\beta}=\frac{\sigma_{\rm v}^2}{kT/\mu m_{\rm p}}$, being $\sigma_{\rm v}$ the mass-weighted mean 
velocity dispersion of dark matter particles, 
and $\tilde{\beta} = \beta[1+f(1/\beta-1)\Omega_{\rm b}/\Omega_{\rm m}]$,  
%
%
where $f$ is the fraction of the barionic matter in the hot gas, and $\Omega_{\rm b}$ is the density parameter of the
baryonic matter.

Combining Eq. (\ref{eq:vifi}), 
Eq. (\ref{eq:psixi}) and Eq. (\ref{eq:conn}) 
I finally obtain:
\begin{eqnarray}
kT&=&1.58\left( \nu +1\right) \frac{\mu }{\beta }\frac{1}{\psi \xi }
\Omega _{o}^{\frac{1}{3}}\left( \frac{M}{10^{15}M_{\odot}h^{-1}}\right) ^{\frac{2}{3}}(1+z_{\rm ta}) 
\nonumber \\
& &\left[ 1+\left( \frac{2^{\frac{10}{3}}\pi ^{\frac{2}{3}}}{3^{\frac{2}{3}}}\right) \psi \xi \rho _{b,ta}^{\frac{2}{3}}\frac{1}{H_{0}^{2}\Omega _{0}M^{\frac{8}{3}} (1+z_{\rm ta})}\int_0^{r_{\rm eff}} \frac{L^{2}}{r^{3}}dr-\frac{2}{3}\frac{\Lambda }{\Omega_0 H_0^{2}(1+z_{\rm ta})^3}\left( \psi \xi \right) ^{3}\right] keV
\label {eq:tem} 
\end{eqnarray}
%
%
%
%
Eq. (\ref{eq:tem}) can be also equivalently written by using Lilje notation, in terms of $r_{\rm vir}$:
\begin{eqnarray}
kT&=&0.94\left( \nu +1\right) \frac{\mu }{\beta }\left(\frac{r_{\rm ta}}{r_{\rm vir}}\right)\left( \frac{\rho _{ta}}{\rho _{b,ta}}\right) ^{\frac{1}{3}}\Omega _{0}^{\frac{1}{3}}\left( \frac{M}{10^{15}M_{\odot}h^{-1}}\right) ^{\frac{2}{3}}(1+z_{\rm ta}) 
\nonumber \\
& & \left[1+\frac{15r_{vir}\rho _{b,ta}}{\pi ^{2}H_0^{2}\Omega_0 \rho_{\rm ta}^{3}r_{ta}^{9}(1+z_{\rm ta})} \int_0^{r_{\rm vir}} \frac{L^2 dr}{r^3}-\frac{2}{3}\frac{\Lambda }{H_0^{2}\Omega_0 }\left( \frac{r_{vir}}{r_{ta}}\right) ^{3} \left(\frac{\rho _{\rm b,ta}}{\rho_{\rm ta} }\right) \frac{1}{(1+z_{\rm ta})^3}\right] keV
%
\label {eq:temp} 
\end{eqnarray}
\begin{figure}
\label{Fig. 1} \centerline{\hbox{
Fig. (3)
\psfig{file=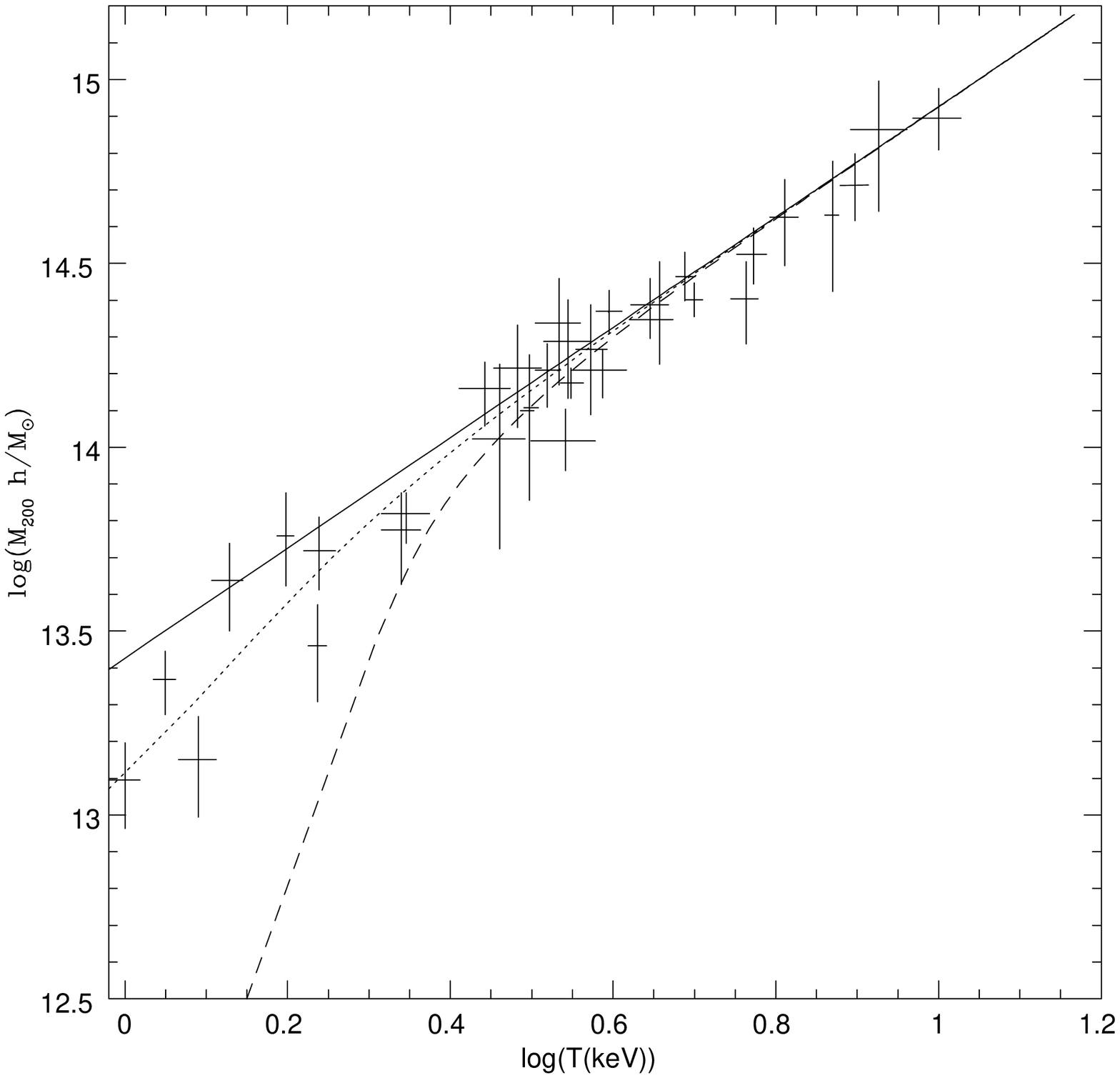,width=9cm} Fig. (4)
\psfig{file=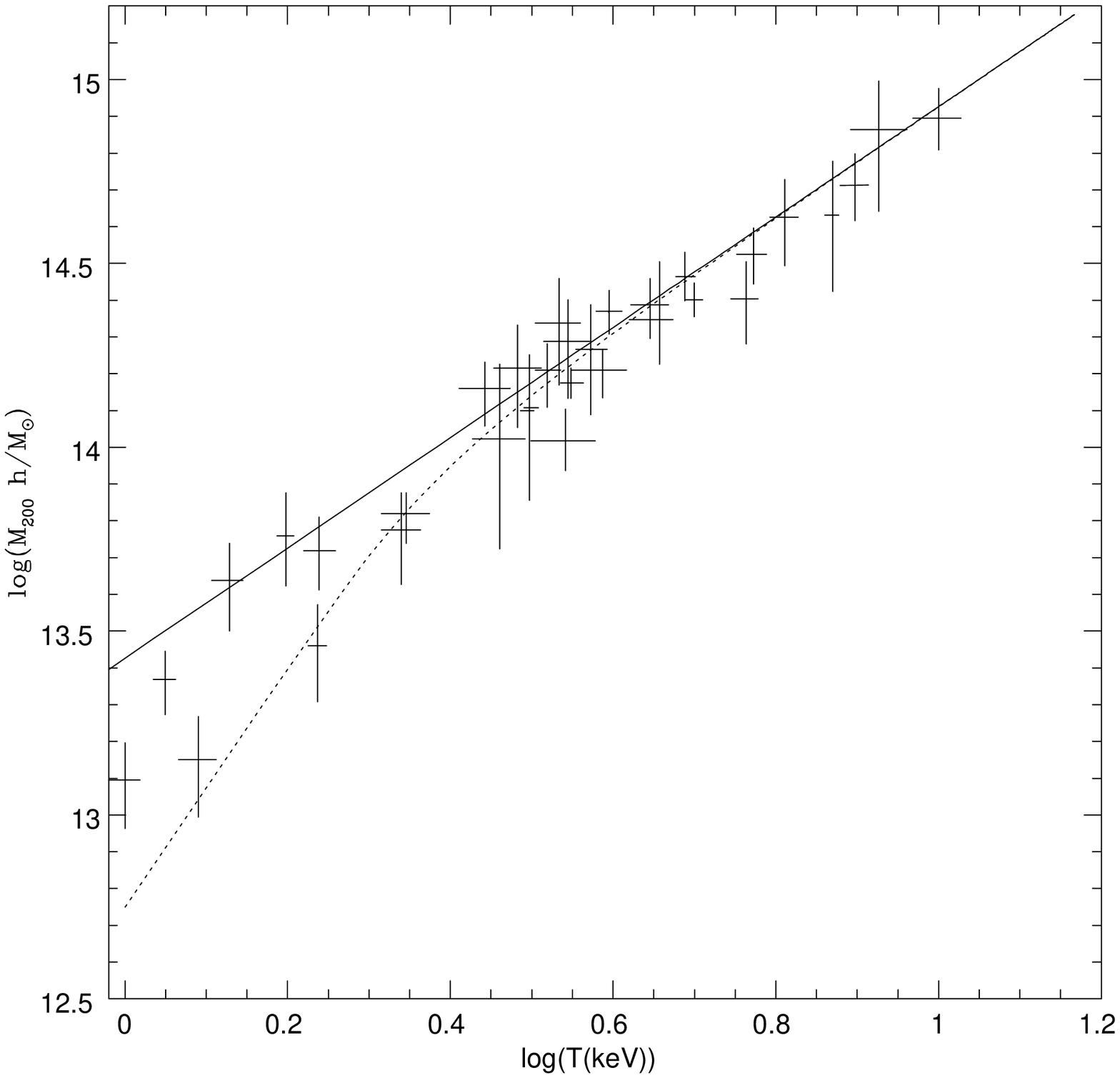,width=9cm}
}}
{\bf Figure 3.} M-T relation predicted by the modified cluster continous-formation model. 
The solid line is the prediction of the quoted model, for $\Omega_{\Lambda}=0$, $\Omega_{\rm 0}=1$, shifted downwards, similarly to AC, to fit the FRB observational data in the massive end. 
The dotted and dashed line represents the prediction of the same model
for $L \neq 0$, $\Omega_{\Lambda}=0$, $\Omega_{\rm 0}=1$ and $L \neq 0$, $\Omega_{\Lambda}=0$ and $\Omega_{\rm 0}=0.3$, respectively.

{\bf Figure 4.} 
Same as Fig. 3, but now the dotted line represents the prediction of the modified cluster continous-formation model
for $L \neq 0$, $\Lambda \neq 0$, ($\Omega_{\Lambda}=0.7$ and $\Omega_{\rm 0}=0.3$).
\end{figure}
%
%

Using $M=\frac{4 \pi}{3} \rho_{\rm b} \Delta_{\rm vir} r_{\rm vir}^3$ and $M=\frac{4 \pi}{3} \rho_{\rm ta} r_{\rm ta}^3$, 
remembering that 
$\rho_{\rm b} (z)=\rho_{0}
\left[\Omega_0 (1+z)^3 +(1-\Omega_0) \right]=\rho_{0} E(z)^2$ 
and $E^2(z)=\frac{\Omega_0 (1+z)^3}{\Omega(z)}$
(Bryan \& Norman 1998),
we see that:
\begin{equation}
 \left(\frac{r_{\rm ta}}{r_{\rm vir}}\right)\left( \frac{\rho _{ta}}{\rho _{b,ta}}\right) ^{\frac{1}{3}}\Omega _{o}^{\frac{1}{3}} (1+z) \propto \left[\frac{\Omega_0}{\Omega(z)}\right]^{1/3} \Delta_{\rm vir}^{1/3} (1+z)
\end{equation}
where
\begin{equation}
\Delta_{\rm vir}(z) \simeq (18 \pi^2+82 y-39 y^2)
\label{eq:over}
\end{equation}
(Bryan \& Norman 1998), being $y \equiv \Omega(z)-1$. These are accurate to 1\% in the range $\Omega(z)=0.1-1$. 
The value of $\Omega(z)$ is:
\begin{equation}
\Omega(z)=\frac{\Omega_0(1+z)^3}{\Omega_0(1+z)^3+
(1-\Omega_0)}
\label{eq:omeg}
\end{equation}
The previous result reduces to the standard one of late-formation approximation (see Eq. 8 of V2000) for $\nu=L=\Lambda=0$. 
%

The value of $r_{eff}$, or the ratio $ \psi=r_{eff}/r_{ta}$ is obtained, similarly to Lahav et al (1991)
by using Eq. (\ref{eq:virial2}) which leads to 
the cubic equation:
\begin{equation}
1-\nu+\left( \xi \psi \right) ^{3}\left[ \eta \nu+2\eta \right] -\psi \left( 2+\eta \xi ^{3}\right) -\frac{27}{32}\frac{\xi ^{9}\psi }{\rho _{ta}^{3}\pi ^{3}Gr_{ta}^{8}}\left( \nu \int_{0}^{r_{eff}}\frac{L^{2}}{r^{3}}dr+\int_{0}^{r_{ta}}\frac{L^{2}}{r^{3}}dr\right)=0
\end{equation}
where 
\begin{equation}
\eta=\frac{\Lambda}{4 \pi G \rho_{\rm ta}}=
\frac{\Lambda r_{\rm ta}^3}{3 G M}= 
\frac{2 \Omega_{\Lambda}}{\Omega_0}
\left(\frac{\rho_{\rm ta}}{\rho_{\rm ta,b}}\right)^{-1}(1+z_{\rm ta})^{-3}
\label{eq:zdep}
\end{equation}
%
%

The previous equation reduces to the standard result $\frac{r_{\rm vir}}{r_{\rm ta}}=1/2$, for $\Lambda=L=\nu=0$, $\Omega_0=1$.
%
%
%
%

The derivation of the previous relation 
is fundamentally based on the approximation of cluster formation with the evolution of a spherical top-hat density perturbation (Peebles 1993) and on the additional assumption that each cluster observed at a redshift $z$ has just reached the moment of virialization. This last assumption is currently known as the late-formation 
approximation, which is a good one in a critical $\Omega_{\rm 0}=1$, because for this value of $\Omega$ massive clusters develop rapidly at all redshifts and the moment of virialization is always close to that of observation. In other terms for $\Omega_{\rm 0}=1$, the accretion rate remains sufficiently high, and this implies that the clusters we actually observe attained their observed masses recently. In the $\Omega_{\rm 0}<1$ case cluster formation is ``shutting down"  and it is necessary to take account of the differences between the moment of virialization and that of observation. The problem becomes worse going through $\Omega_{\rm 0}<<1$: in fact in the late-formation approximation $M_{\rm vir}$ rises steadily  since $\rho_{\rm b} \Delta_{\rm vir}$ declines indefinitely, while we expect that the cluster formation is going to stop. 
As shown by V2000, to obtain the proper normalization and time evolution of the M-T relation, one has to account:\\
a) for the continuous accretion of mass of clusters;\\
b) for the non-zero density at $r_{\rm vir}$, requiring a change in the virial theorem by including a surface pressure term.  

The M-T relation derived by means of a model of continuous accretion, as that of V2000, differs from the late-formation model in both normalization and time-dependent behavior. A comparison of the normalization predicted by the late-formation model with that predicted by simulations of Evrard, Metzler \& Navarro (1996) shows that when $\Omega_0=1$ this normalization is only 4\% below the empirical value, but it lies 20\% below it for $\Omega_0=0.2$.
In the case of V2000 model and for a power-law spectrum, a comparison with the same simulations show that the temperature normalization of the $n=-2$  case deviates by less than 10\% over the range $0.2<\Omega_{\rm 0}<1$ and by $\simeq 18\%$ in the case $n=-1$ (V2000). The normalization obtained by the V2000 model, even if it is more accurate than that given by the late formation, or that by AC which is in agreement with hydro-simulations, show a noteworthy discrepancy when compared with X-ray mass estimates (about 50\% for the AC model; see also V2000). 
One possible source for differences in theoretical and observational normalizations may be due to the fact that $\beta$ is different in the two cases because of systematic selection effects. For example, as shown by Bryan \& Norman (1998), increasing the resolution of simulations there is an increase in the value of $\beta$.
So summarizing, for what concerns normalization,
the continuous formation model gives more precise results than the late formation one, but in any case if we want to fit observations we need to shift the normalization (see AC). In any case, we have to stress that the continuous formation model must be considered as a noteworthy improvement on the late formation and is physically more consistent than the THM:
as stressed by V2000, the top-hat model does not account properly for energy and mass accumulation during the early stages of cluster formation or the confining effects of matter that continues to fall in, both of which increase the temperature associated with a given mass (note that this last problem has been removed in the THM model). 

For the purpose of studying cluster evolution, the change in the M-T relation with $z$ is more important than its normalization: although the differences in evolution of the M-T relation are of the order of $10\%$ (V98), when it is introduced in the Press-Schechter model, they are amplified by the exponential term in the Press-Schechter formula.   

In the next section, I'll extend the V2000 model to take account of tidal interaction between clusters and the evolution of the M-T relation, comparing the results with the those of the top-hat model.

\section{Revisiting the continuous formation model}

In the previous section, I obtained the M-T relation for a universe with $\Lambda \neq 0$ and $ L \neq 0$ in the late-formation approximation. Some shortcomings of this approach has been summarized in the previous section (see also Viana \& Liddle 1996; Kitayama \& Suto 1996; Eke et al. 1996). The late-formation approximation is a good one for many purposes, but a better one can be obtained in the low-$\Omega$ limit. As can be found in the literature, there are two ways of improving the quoted model. One is to define a formation redshift $z_{\rm f}$ at which a cluster virializes and after the properties of observed clusters at $z$ are obtained by integrating over the appropriate distribution of formation redshifts (Kitayama \& Suto 1996; Viana \& Liddle 1996). The second possibility is that described by V98, V2000. In this approach, the top-hat cluster formation model is substituted by a model of cluster formation from spherically symmetric perturbations with negative radial density gradients. The fact that clusters form gradually, and not instantaneously, is taken into account in the merging-halo formalism of Lacey \& Cole (1993). In hierarchical models for structure formation, the growth of the largest clusters is quasi-continuous since these large objects are so rare that they almost never merge with another cluster of similar size (Lacey \& Cole 1993). So, Lacey \& Cole (1993) approach extends the Press-Schechter formalism by considering how clusters grow via accretion of smaller virialized objects. As shown by V2000, the mass grows like $M \propto \omega^{-3/(n+3)}$ (Lacey \& Cole 1993; V98; V2000). The virial energy of the cluster, $-E$, can be calculated by integrating the specific energy $\epsilon$, $E=-\int \epsilon dM$. The cluster temperature can be obtained remembering that it is proportional to $\frac{E}{M}$. Integrating Eq.(\ref{eq:coll1}), I get:
\begin{equation}
t=\int\frac{dr}{\sqrt{2\left[ \epsilon+\frac{GM}r+\int_{r_{\rm i}}^{r} \frac{L^2}{M^2r^3}dr+\frac{\Lambda}{6}r^2\right] }}
\label{eq:tmppp}
\end{equation}
A particular shell will collapse if: 
\begin{equation}
\epsilon+\frac{GM}{r_0}+\int_{r_{\rm i}}^{r_0} \frac{L^2}{M^2r^3}dr+\frac{\Lambda}{6}r_0^2=0
\end{equation}
(see V2000),
and the shell reaches its maximum radius (turn-around radius, $r_{\rm ta}$) at a time:
\begin{equation}
t_{\rm ta}=\int_{0}^{r_{\rm ta}} \frac{dr}{\sqrt{2\left[ \epsilon+\frac{GM}r+\int_{r_{\rm i}}^{r} \frac{L^2}{M^2r^3}dr+\frac{\Lambda}{6}r^2\right]}}
\label{eq:tmp}
\end{equation}
Eq. (\ref{eq:tmp}) can be written in an equivalent form (see Del Popolo \& Gambera 1998, 1999; Bartlett \& Silk 1993), as:
\begin{eqnarray}
t_{ta}&=&\int_{0}^{r_{ta}}\frac{dr}{\sqrt{2\left[ GM\left( \frac{1}{r}-\frac{1}{r_{ta}}\right) +\int_{r_{ta}}^{r}\frac{L^{2}}{M^{2}r^{3}}dr+\frac{\Lambda}{6}(r^2-r^2_{\rm ta})\right] }}=
\nonumber \\
& &
2H_0^{-1}\Omega_0 ^{-1/2}\xi ^{3/2}\int_{0}^{1}
\frac{y^{2}dy}{\sqrt{1+
\left[ \frac{9\xi ^{9}}{4\pi ^{2}\rho ^{2}r_{ta}^{10}\Omega H^{2}}\int_{1}^{y^{2}r_{ta}}\frac{L^{2}}{y^{5}}dy\right]y^2
+\left( \frac{\Omega _{\Lambda }}{\Omega _{0}}\right) \xi ^{3}y^{6}-\left[ 1+\left( \frac{\Omega _{\Lambda }}{\Omega _{0}}\right) \xi ^{3}\right] y^{2}}}
\label{eq:tmp1}
\end{eqnarray}
%
%
%
\begin{figure}
\label{Fig. 1} \centerline{\hbox{
Fig. (5a)
\psfig{file=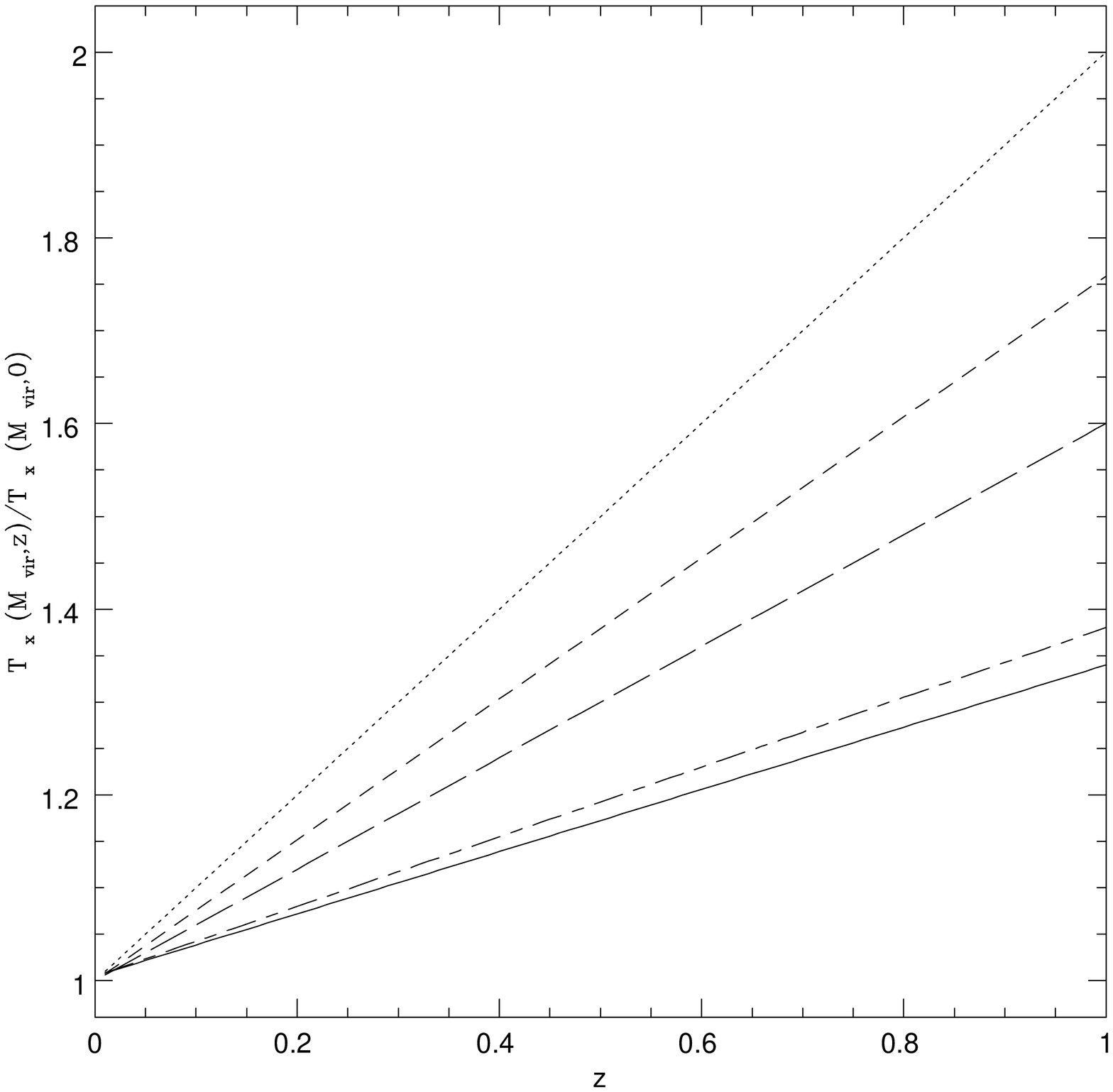,width=9cm} Fig. (5b)
\psfig{file=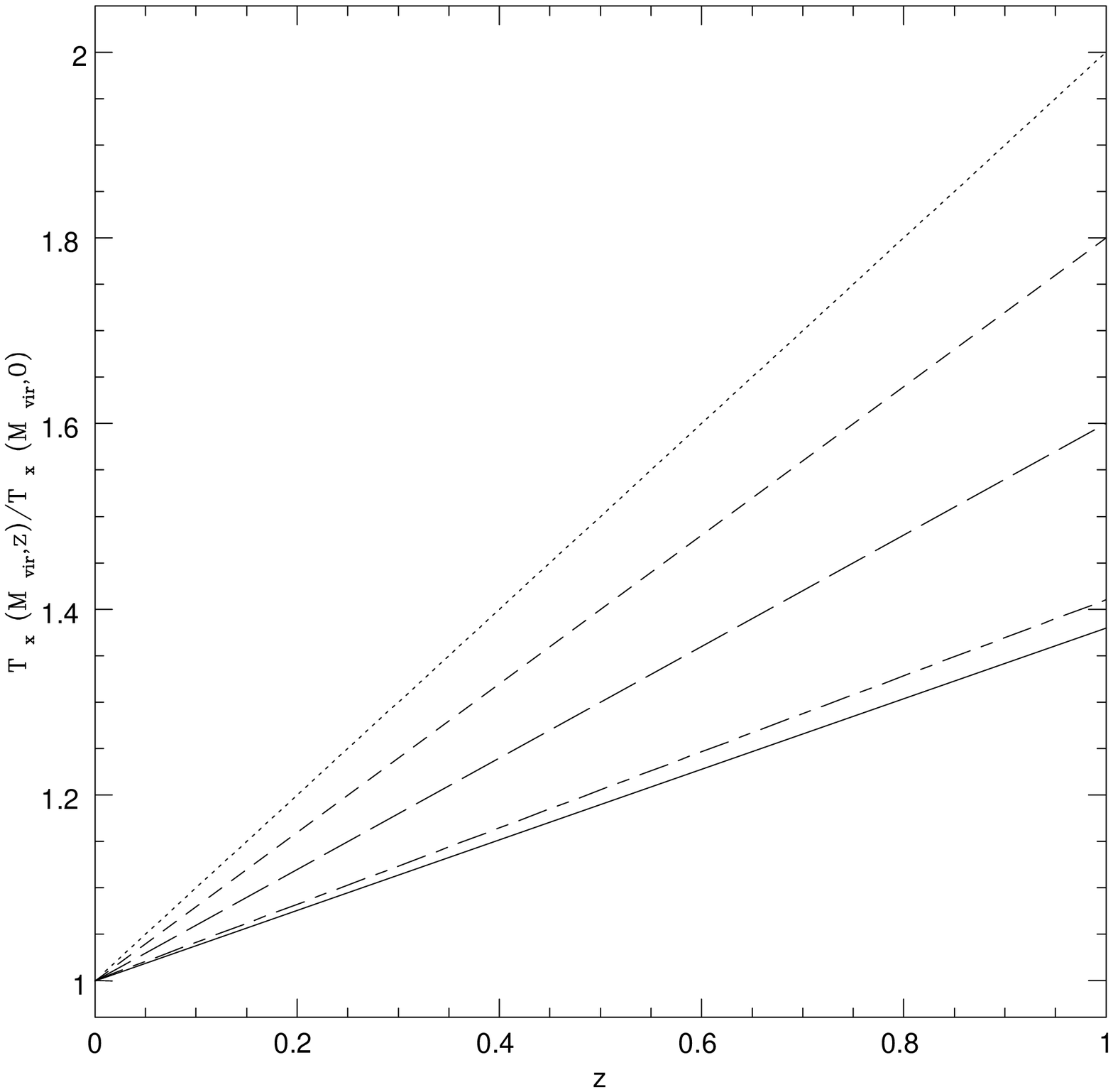,width=9cm}
}}
{\bf Figure 5a.} Temperature evolution predicted by the top-hat M-T relation. 
The dotted line represents the prediction for $\Omega_0=1$ and $L=0$ which coincides with the ``classical" prediction, $T_{\rm X} \propto (1+z)$. The short-dashed line represents 
the case $\Omega_0=0.2$ and $L=0$, coinciding with the late formation model described in V2000 (Eq. 8). 
The long-dashed line represents $\Omega_0=1$ and $L \neq 0$. 
Finally, the short-long-dashed line represents the case $L \neq 0$, $\Lambda \neq 0$ ($\Omega_0=0.2$, $\Omega_{\Lambda}=0.8$) and
the solid line the case $L \neq 0$, $\Lambda=0$, $\Omega_0=0.2$. 

{\bf Figure 5b.} Same as Fig. 5a but now  
the short-dashed line represents 
the case $\Omega_0=0.3$ and $L=0$, while  
the short-long-dashed line represents the case $L \neq 0$, $\Lambda \neq 0$ ($\Omega_0=0.3$, $\Omega_{\Lambda}=0.7$) and
the solid line the case $L \neq 0$, $\Lambda=0$, $\Omega_0=0.3$. 
\label{Fig. 5}
\end{figure}
The shell collapse at $t_{\rm c}=2 t_{\rm ta}$.
In the case $L=0$, Eq. (\ref{eq:tmp}) can be analytically integrated and inverted to get $\epsilon(t)$:
\begin{equation}
\epsilon(t)=\frac12 \left(\frac{2 \pi G M}{t}\right)^{2/3}=
\frac12 \left(\frac{2 \pi G M}{t_{\Omega}}\right)^{2/3} \left(\frac{t_{\Omega}}{t}\right)^{2/3}=
\frac12 \left(\frac{2 \pi G M}{t_{\Omega}}\right)^{2/3} (x-1)
\label{eq:eps}
\end{equation}
where we have defined $t_\Omega =\frac{\pi \Omega _0}{H_o\left( 1-\Omega _0-\Omega _\Lambda \right) ^{\frac 32}}$ and $x=1+(\frac{t_{\Omega}}{t})^{2/3}$ which is connected to mass by $M=M_{\rm 0} x^{-3 m/5}$ (V2000). 
\begin{figure}
\label{Fig. 1} \centerline{\hbox{
Fig. (6a)
\psfig{file=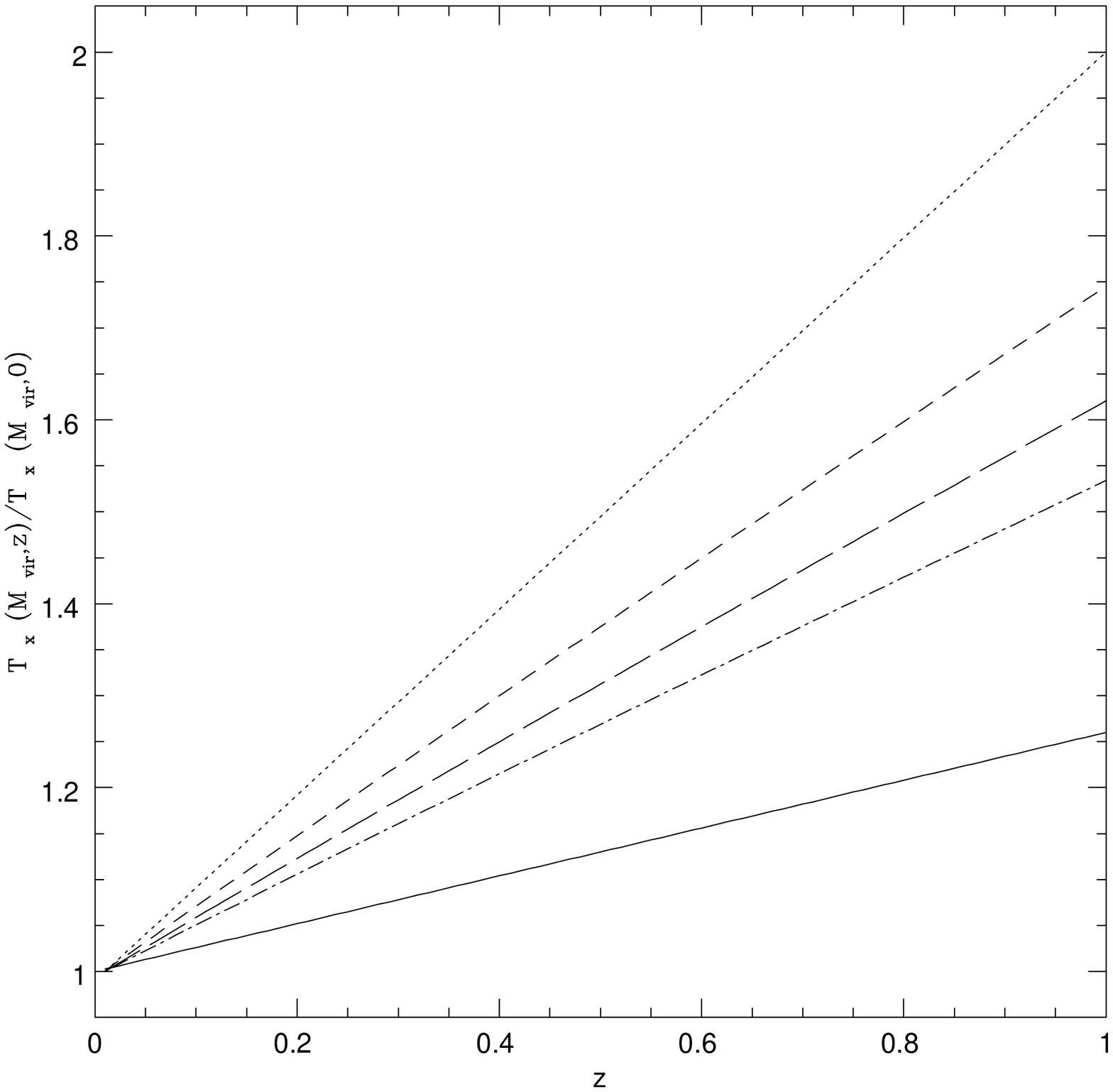,width=9cm}
\psfig{file=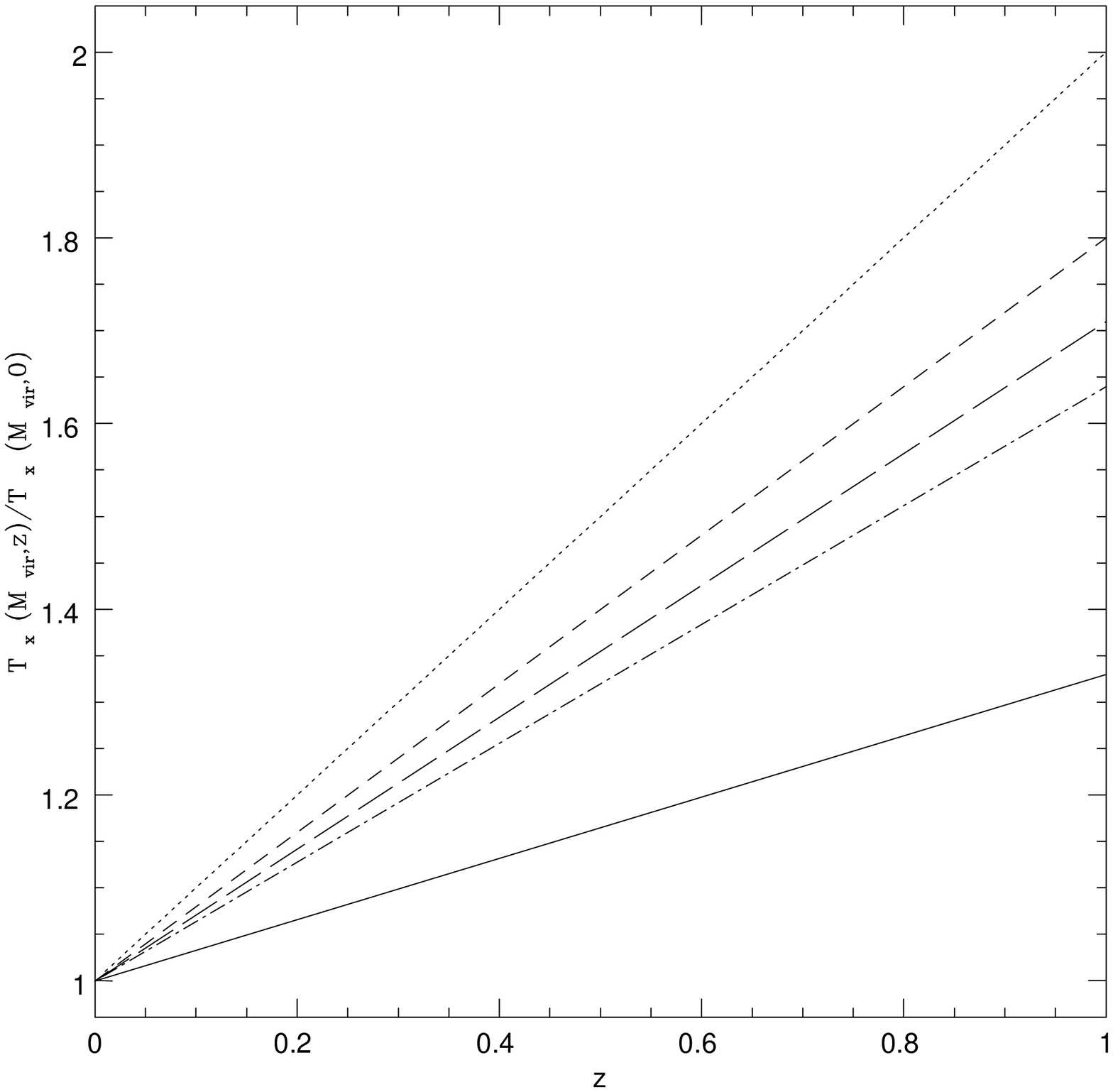,width=9cm}
}}
{\bf Figure 6a.} Temperature evolution predicted by the modified cluster continuous-formation model in the case $\Omega_0=0.2$.
The dotted line represents the ``classical'' prediction, $T_{\rm X} \propto (1+z)$. 
The short-dashed line represents the late-formation approximation as expressed by Eq. (8) in V2000, namely $T_{\rm X} \propto \Delta_{\rm vir}^{1/3} \left[\frac{\Omega_{\rm 0}}{\Omega_{\rm 0}(z)}\right]^{1/3} (1+z)$.
The long-dashed line and the dot-dashed one plot Eq. (\ref{eq:kT}) for $L=0$ and $n=-2$, $n=-1$, respectively. 
The solid line plot the same equation for $L \neq 0$ and $n=-1$. 
%
%

{\bf Figure 6b.} Same as Fig. 6a but for $\Omega_0=0.3$.
\end{figure}

Integrating with respect to mass $\epsilon(t)$, and diving by mass, one gets, as shown by V2000:
\begin{equation}
\frac{E}{M}=-\frac{\int \epsilon dM}{M}=
\frac{3m}{10(m-1)}\left( \frac{2\pi G}{t_\Omega }\right) ^{\frac 23}M^{\frac 23}\left[ \frac 1m+\left( \frac{t_\Omega }t\right) ^{\frac 23}\right]
\end{equation}
In the case $L \neq 0$, it is numerically possible to integrate and invert Eq. (\ref{eq:tmp}) finding the specific energy,
which we can indicate with $\epsilon_{\rm L}(t)$ and finally:
\begin{equation}
\frac{E}{M}=-\frac35 \frac{m}{m-1}\epsilon_{\rm L}(t) 
\end{equation}
An approximate relation for $\epsilon_{\rm L}(t)$, in an Einstein-de Sitter Universe can be obtained as follows.
It is possible to obtain the turn-around radius, $r_{\rm ta}$ by solving Eq. (\ref{eq:tmp1}) (with $\Lambda=0$) for a given mass and a given epoch of interest. This is related to the binding energy of the shell enclosing the mass $M$ by Eq. (\ref{eq:coll1}) with $\dot r=0$. 
In turn, the binding energy is uniquely given by the linear overdensity $\delta_{\rm i}$ at some arbitrary early time.
We may now use the linear theory to ``propagate" the overdensity to the ``chosen" time to find the linear overdensity at turn-around, $\delta_{\rm c}$.
Using the relation between $v$ and $\delta_{\rm i}$ for the growing mode (Peebles 1980) in Eq. (\ref{eq:coll1}), with $\Lambda=0$, at an early time, it is possible to connect $\epsilon$ and $\delta_c$ (see Bartlett \& Silk 1993)):
\begin{equation}
\epsilon=\frac56 \Omega_0 H_0^2 x_1^2(\delta_{\rm i}/a_{\rm i})=\frac12 \left(
\frac{2 \pi G M}{t}
\right)^{2/3}
\label{eq:epsil}
\end{equation}
Using the formula for $\delta_{\rm c}$ given in Del Popolo \& Gambera (1999, Eq. 14), Del Popolo \& Gambera (2000, Eq. 6), is possible to write:
\begin{equation}
\epsilon_{\rm L}=-\frac12 \left(\frac{2 \pi G M}{t}
\right)^{2/3}\delta_{\rm c}
%
%
=
\frac12 \left(\frac{2 \pi G M}{t}\right)^{2/3}
\left[
1+\frac{r_{\rm ta}}{G M^3} \int_0^{r} \frac{L^2 dr}{r^3}\right]
\label{eq:epsil1}
\end{equation}
which reduces to Eq. (\ref{eq:epsil}) when $L \rightarrow 0$.

Eq. (\ref{eq:epsil1}) can also be written as:
\begin{equation}
\epsilon_{\rm L}=-\frac12 \left(\frac{2 \pi G M}{t_{\Omega}}\right)^{2/3}(x-1)
\left[
1+\frac{2^{7/3} \pi^{2/3} \xi \rho_{\rm b}^{2/3}}{3^{2/3} H^2 \Omega M^{8/3}} \frac{1}{x-1}
\int_0^{r} \frac{L^2 dr}{r^3}\right]
\label{eq:ener1}
\end{equation}
Defining $M=M_{\rm 0} x^{-3 m/5}$ and 
\begin{equation}
F=\frac{2^{7/3} \pi^{2/3} \xi \rho_{\rm b}^{2/3}}{3^{2/3} H^2 \Omega} 
\int_0^r \frac{L^2 dr}{r^3}
\end{equation}
Eq. (\ref{eq:ener1}) becomes:
\begin{equation}
\epsilon_{\rm L}=-\frac12 \left( \frac{2 \pi G}{t_{\Omega}}\right)^{\frac 23}M^{2/3}
\left[ \left( \frac M{M_o}\right) ^{-\frac 5{3m}}-1\right] \left[ 1+\frac F{M^{\frac 83}}\frac 1{\left( \frac M{M_o}\right) ^{-\frac 5{3m}}-1}\right]  
\label{eq:ener2}
\end{equation}
Integrating with respect mass, and dividing again by $M$, I get:
\begin{equation}
\frac{E}{M}=-\frac{\int \epsilon dM}{M}=
\frac{3m}{10(m-1)}\left( \frac{2\pi G}{t_\Omega }\right) ^{\frac 23}M^{\frac 23}
\left[
\frac 1m+\left( \frac{t_\Omega }t\right) ^{\frac 23}
+\frac{K(m,x)}{M^{8/3}}
\right]
\label{eq:em} 
\end{equation}
\begin{eqnarray}
K(m,x)&=&\left (m-1\right )F x {\it LerchPhi}
(x,1,3m/5+1)-
\nonumber \\
& &
\left (m-1\right )F{\it LerchPhi}(x,1,3m/5)
\end{eqnarray}
where the {\it LerchPhi} function is defined as follows: 
\begin{equation} 
LerchPhi(z,a,v)=\sum_{n=0}^{\infty} \frac{z^n}{(v+n)^a}
\end{equation}
%
\footnote{This definition is valid for $|z| < 1$. By analytic continuation, it is extended to 
the whole complex z-plane, for each value of a.
If the coefficients of the series representation of a hypergeometric function are rational functions of the summation indices, then the hypergeometric function can be expressed as a linear sum of Lerch Phi functions. 
Reference: A. Erdelyi, 1953, Higher Transcendetal Functions, Volume 1, chapter 1, section 11.} 
%
If $K=0$, Eq. (\ref{eq:em}) reduces to Eq. (10) of V2000. As stressed by V2000, some factors give rise to an higher value of $E/M$ with respect the case of the late-formation value. The $m/(m-1)$ value which accounts for the effect of early infall. The $1/m$ value in the square bracket of Eq. (\ref{eq:em}) which accounts for the cessation of cluster formation when $t>>t_{\rm \Omega}$. Finally in Eq. (\ref{eq:em}) a new term is present, which comes from the tidal interaction.
In order to obtain an expression for the kinetic energy, starting from $E/M$, I use the virial theorem with the surface pressure term correction as in V2000 and Eq. (\ref {eq:conn}). In this way, I finally get: 
\begin{equation}
kT=\frac 25a\frac{\mu m_p}{2\beta} \frac m{m-1}\left( \frac{2\pi G}{t_\Omega }\right) ^{\frac 23}M^{\frac 23}
\left[
\frac 1m+\left( \frac{t_\Omega }t\right) ^{\frac 23}
+\frac{K(m,x)}{M_0^{8/3}}
\right]
\label{eq:kT}
\end{equation}
where $a=\frac{\overline{\rho}}{2 \rho(r_{\rm vir})-\overline{\rho}}$ is the ratio between kinetic and total energy (V2000).
Using the relation $\Delta_{\rm vir}=\frac{8 \pi^2}{H t^2}$ (see V2000), and in the early-time limit: ($t<<t_{\Omega}$), Eq. (\ref{eq:kT}), reduces to:
\begin{equation}
kT=\frac 25 \frac m{m-1}a\frac{\mu m_p}{2\beta}  G M^{\frac 23}
\left(
\frac{4 \pi}{3} \rho_{\rm b} \Delta_{\rm vir}
\right)^{1/3}
\label{eq:kT1}
\end{equation}
which, in the case $n \sim -2$, $a \sim 2$ is identical 
to the late-formation formula, described in V2000 (see their Eq. (8)).
Normalizing Eq. (\ref{eq:kT}) similarly to V2000, I get:
\begin{equation}
kT \simeq 8 keV \left(\frac{M^{\frac 23}}{10^{15}h^{-1} M_{\odot}}\right)
\frac{
\left[
\frac 1m+\left( \frac{t_\Omega }t\right) ^{\frac 23}
+\frac{K(m,x)}{M^{8/3}}
\right]
}
{
\left[
\frac 1m+\left( \frac{t_\Omega }{t_{0}}\right) ^{\frac 23}
 +\frac{K_0(m,x)}{M_0^{8/3}}
\right]
}
\label{eq:kT1}
\end{equation}
where $K_0(m,x)$ indicates that $K(m,x)$ must be calculated assuming $t=t_0$

Eq. (\ref{eq:kT1}) when compared to the result of V2000 (Eq. 17) shows an additional term, mass dependent. This means that, as in the case of the top-hat model, the M-T relation is no longer self-similar showing a break at the low mass end (see next section).

\section{Results}

The results of the models described in the previous sections are plotted in Fig. 1-6.

Fig. 1 compares the M-T relation obtained by means of the THM model with the FRB data. The solid line is the prediction of 
Eq. (\ref{eq:temp}), for $\Omega_{\Lambda}=0$, $\Omega_{\rm 0}=1$, shifted downwards, similarly to AC, to fit the observational data in the massive end. The reason of the shift is due to the fact that, as reported in the introduction, X-ray mass estimates lead to normalizations about 50\% higher than hydro-simulations, (while mass estimates from galaxy velocity dispersion is consistent with simulation results (HMS)). The coincidence of the slope of the best fit to data and the theoretical model indicates that the process that the theoretical models does not take properly into account is  probably happening at small scales (AC). 
The dotted line represents the prediction of Eq. (\ref{eq:temp}) for $L \neq 0$, $\Omega_{\Lambda}=0$, $\Omega_{\rm 0}=1$. The plot shows a non-self-similar behavior. A break at $T_{\rm X}\sim 3 {\rm keV}$ is evident and in this low mass range the power index $\alpha \sim 1.8$. It is interesting to note that the mass estimates using the observational $\beta$-model lead to a power index $1.7<\alpha<1.8$. Originally this behavior was interpreted as an artifact of the $\beta$-model (HMS), but the observation of a similar behavior for masses estimated from the resolved temperature profile (FRB) and confirmed by other studies (e.g., Xu, Jin \& Wu (2001)), who found the break at $T_{\rm X}=3-4 {\rm keV}$. 
The quoted bent is interpreted in different ways.
FRB suggests that it is due to different formation redshifts, but this conclusion is based upon 
the assumption that the temperature is constant after the formation time. This assumption is probably not true since clusters continue to accrete mass in time, even after cluster has formed. So as argued by Mathiesen (2001), using simulations results, the formation time might not be so important. 
Another interpretation is based on the assumption that the cluster medium is preheated in the early phase of cluster formation (see Xu, Jin \& Wu 2001).   
In the AC study, the quoted break was reproduced (justified), based on the scatter in the density field for a low density $\Lambda$CDM cosmology. The AC model is not able to distinguish between the effect of formation redshift from scatter in the initial energy of the cluster or its initial non-sphericity. In any case, non-sphericity introduces an asymmetric scatter dependent on mass for the M-T relation altering its slope at $T_{\rm X} \sim 3 {\rm keV}$. 
As stressed by AC, the calculation of the effects of the non-spherical shape of the initial proto-cluster are not very rigorous and should be considered as an estimate of the actual corrections.

In the model of this paper, the bent is entirely justified in terms of cluster 
tidal interaction with the neighboring ones, or in other terms it is strictly connected to the asphericity of clusters (see Del Popolo \& Gambera 1999 for a discussion on the relation between angular momentum acquisition, asphericity and structure formation). 
Non-sphericity introduces an asymmetric bent, dependent on mass, in the M-T relation that gives rise to a different slope at the low mass end ($T \sim 3 {\rm keV}$): the lower the mass the larger the bent. 
A possible explanation for the reason of the origin of the bent is the following. 
The M-T relations found in this paper differ from previous models (e.g. V2000, AC),
for the presence of a term
depending on mass and angular momentum, $L$,  
originating from the gravitational interaction of the quadrupole moment
of the protocluster with the tidal field of the matter of the
neighboring protostructures.
This last term changes the dependence of the temperature on the mass, $M$, in the M-T relation. 
The origin of the quoted term is due to: \\
a) the change in the energetics of the
collapse model and therefore of the 
turnaround epoch and the value of the
threshold parameter $\delta_{c}$ that now is a function of the mass,$M$,
produced by the introduction of another potential energy term
($\frac{L(r)^2}{M^2 r^3}$) in
$Eq.(\ref{eq:coll})$.  
Angular momentum acquisition influence the value of $\delta _{\rm c}$
which increases for peaks of low mass  and remains unchanged for
high mass peaks. 
Similarly to what is shown in
Del Popolo \& Gambera (1998, 1999), the threshold, $\delta _{\rm c}$, is a
decreasing function of the mass, $M$. 
Since temperature, $T$, is directly proportional to the specific energy, $\epsilon$, of infalling matter and this in turn is directly proportional to the threshold parameter, $\delta _{\rm c}$, (see V2000), we have as a result that less massive clusters must be hotter while clusters in the high mass end have their temperature unchanged.
%
\\
b) the modification of the M-T relation produced
by the alteration of the partition of energy in virial equilibrium. 
Moreover the partition of energy in virial
equilibrium is altered and consequently also the M-T relation is modified.

Coming back to Fig. 1, the dashed line represents the prediction of Eq. (\ref{eq:temp}) for $L \neq 0$, $\Omega_{\Lambda}=0$ and $\Omega_{\rm 0}=0.3$.  Fig. 1 shows that in an Einstein-de Sitter universe the bent is under-estimated, while in a OCDM model with $\Omega_{\rm 0}=0.3$ it is overestimated. 
The comparison shows that the FRB data is 
able to rule out very low $\Omega_0$ models ($<0.3$). 

Fig. 2 represents the same calculation of the previous figure but now $\Omega_{\Lambda} \neq 0$. The solid line is equivalent to the solid one in Fig. 1. 
The dashed line represents the case $L \neq 0$, $\Lambda \neq 0$, ($\Omega_{\rm 0}=0.3$, $\Omega_{\Lambda}=0.7$). 
It is evident that the effect of a non-zero cosmological constant is that of reducing the effect of $L$, and that the $\Lambda$CDM cosmology, with $\Omega_{\rm 0}=0.3$, $\Omega_{\Lambda}=0.7$, is in better agreement with the observed bent than the $\Omega_{\rm 0}=0.3$ OCDM model.

%
%

The M-T relation predicted by the modified continous formation model is plotted in Fig. 3 and Fig. 4.
In Fig. 3, the solid line is the prediction of the quoted model, for $\Omega_{\Lambda}=0$, $\Omega_{\rm 0}=1$, shifted downwards,  similarly to AC and to the top-hat model (Fig. 1-2), to fit the FRB observational data in the massive end. 
The dotted and dashed line represents the prediction of the same model for $L \neq 0$, $\Omega_{\Lambda}=0$, $\Omega_{\rm 0}=1$ and $L \neq 0$, $\Omega_{\Lambda}=0$ and $\Omega_{\rm 0}=0.3$, respectively. 
In agreement with the predictions of the top-hat model, the M-T relation shows a break at the low mass end ($T \simeq 3 {\rm keV}$). The break is less ``evident" in the case of a flat universe while much more evident in the case of an open one.  
A comparison with the FRB data shows,  
similarly to the top-hat model prediction, that in an Einstein-de Sitter universe the bent is under-estimated, while in a OCDM model with $\Omega_{\rm 0}=0.3$ it is overestimated. 
It is to be noted that the modified continuous formation model predicts a slightly larger bent than the top-hat model. 

Fig. 4 plots the M-T relation for the modified continuous formation model in a $\Lambda$CDM model. In this case, 
$\epsilon_{\rm L}$ and then the M-T relation is calculated numerically by 
integrating and inverting Eq. (\ref{eq:tmppp}).
The solid line is the prediction for $\Omega_{\Lambda}=0$, $\Omega_{\rm 0}=1$, shifted downwards, as previously described and motivated, while the dotted line represents the prediction of the same model
for $L \neq 0$, $\Omega_{\Lambda}=0.7$ and $\Omega_{\rm 0}=0.3$. The plot shows, similarly to the top-hat model, that the effect of a non-zero cosmological constant is that of reducing the effect of $L$, and that the $\Lambda$CDM cosmology (with 
$\Omega_{\Lambda}=0.7$ and $\Omega_{\rm 0}=0.3$) 
is in better agreement with the observed bent than the $\Omega_{\rm 0}=0.3$ OCDM model.
The $\Lambda$CDM model predicts a slightly larger bent in the M-T relation when compared with the top-hat model.

A comparison between the Fig.1-2 and Fig. 3-4 shows that the prediction of the THM model and the VM model differs only for $1-2\%$. This little difference is possible because we have chosen the same normalization for the two models.

The study of the evolution of the M-T relation predicted by the modified top-hat model, for a fixed value of $M_{\rm vir}$, is plotted in Fig. 5a-b. 

In Fig. 5a, the dotted line represents the prediction for $\Omega=1$ and $L=0$ which coincides with the ``classical" prediction, $T_{\rm X} \propto (1+z)$. 
The short-dashed line represents 
the case $\Omega_{\rm 0}=0.2$ and $L=0$, which  coincides with $T_{\rm X} \propto \Delta_{\rm vir}^{1/3} \left[\frac{\Omega_{\rm 0}}{\Omega_{\rm 0}(z)}\right]^{1/3} (1+z)$, given in V2000 (Eq. 8). As known, (see V98), decreasing $\Omega$ slows down the evolution. 
The long-dashed line represents $\Omega_{\rm 0}=1$ and $L \neq 0$. Comparing this result to the first one $\Omega_{\rm 0}=1$ and $L=0$ 
shows that the effect of $L$ in slowing down the evolution of $T_{\rm X}$ is larger than reducing $\Omega_{\rm 0}$ of a noteworthy value (0.8, see short-dashed line). 
Finally, the short-long-dashed line represents the case $L \neq 0$, $\Lambda \neq 0$, $\Omega=0.2$, $\Omega_{\Lambda}=0.8$ and 
the solid line the case $L \neq 0$, $\Lambda=0$ $\Omega_0=0.2$. 
As expected, reducing the value of $\Omega$ and taking account of tidal interaction of clusters ($\L \neq 0$) produces a larger effect in the slowing down of evolution, while the effect of a non-zero cosmological constant has an opposite effect to that of $L$. Fig. 5b plots the same calculations of Fig. 5a but for the case $\Omega_0=0.3$. As expected, an increase in $\Omega$ produce a more rapid evolution of $T_{\rm X}$.

Fig. 6a, is based upon the second model (modified continuous formation) and calculates the evolution for $\Omega_0=0.2$. The dotted line represents the ``classical'' prediction, $T_{\rm X} \propto (1+z)$. This relation predicts the largest evolution for a given $M_{\rm vir}$.
The short-dashed line represents the late-formation approximation as expressed by Eq. (8) in V2000, namely $T_{\rm X} \propto \Delta_{\rm vir}^{1/3} \left[\frac{\Omega_{\rm 0}}{\Omega_{\rm 0}(z)}\right]^{1/3} (1+z)$.
%
%
In this case, temperature evolution is more modest than the previous one.  
The long-dashed line and the dot-dashed one, plot Eq. (\ref{eq:kT}) for $L=0$ and $n=-2$, $n=-1$, respectively. 
This last two lines show that, (see also V98), the continous formation model produces an even more modest evolution of temperature with respect to the late-formation approximation.
The solid line plot Eq. (\ref{eq:kT}) for $L \neq 0$ and $n=-1$.
%
%
The effect of the angular momentum is that of furtherly increase the slowing down effect of the temperature evolution. 
%
%
Fig. 6b, plots the same calculations of Fig. 6a for $\Omega_{\rm 0}=0.3$.

A comparison between the Fig. 5 and Fig. 6 shows that the prediction, for what concerns the time evolution, of the THM model and the VM model differs $ \leq 10\%$ (for $L \neq 0$, $\Lambda=0$, $\Omega_0=0.2$).

\section{Conclusions}

In this paper, I used two different models in order to study the M-T relation and its time evolution. The first one is based on a modification of the top-hat model in order to take account of an external pressure term in the virial theorem and of angular momentum acquisition by protostructures (see Del Popolo \& Gambera 1999), the second one is a modification of the continuous formation model of V2000, again to take account of angular momentum acquisition by protostructures. 

The main results of the paper can be summarized as follows: \\
1) the effect of angular momentum  acquisition by protostructures is two fold:\\
a) The M-T relation is no longer self-similar: a break in the low mass end ($T \sim 3-4 {\rm keV}$) of the M-T relation is present. The behavior of the M-T relation is as usual, $M \propto T^{3/2}$, at the high mass end, and $M \propto T^{\gamma}$, with a value of $\gamma>3/2$ in dependence of the chosen cosmology. Larger values of $\gamma$ are related to open cosmologies, while $\Lambda$CDM cosmologies give results of the slope intermediate between the flat case and the open case. The FRB data are best fitted by 
$\Lambda$CDM models and by CDM models with $\Omega_0>0.3$.\\
2) The evolution of the M-T relation is more rapid in models with $L=0$. The effect of a non-zero cosmological constant is that of slightly increase the evolution of the M-T relation with respect open models with the same value of $\Omega_0$
This is due to the fact that in a flat $\Omega_{\Lambda}>0$ universe, the universe's density remains close to the critical value later in time, promoting perturbation growth at lower redshift.
The evolution is more rapid for larger values (in absolute value) of the spectral index, $n$.
%
%
\\ 
3) The top-hat model gives comparable results to the continuous formation model, in the range of $z$ and $\Omega_0$ considered. For sake of precision, the top-hat model has  a lower value of normalization. A comparison of the normalization predicted by the late-formation model with that predicted by simulations of (Evrard, Metzler \& Navarro 1996) shows that when $\Omega_0=1$ this normalization is only 4\% below the empirical value, but it lies 20\% below it for $\Omega_0=0.2$.
In the case of V2000 model and for a power-law spectrum, a comparison with the same simulations show that the temperature normalization of the $n=-2$  case deviates by less than 10\% over the range $0.2<\Omega_{\rm 0}<1$ and by $\simeq 18\%$ in the case $n=-1$ (V2000).
Given the same value of normalization for both models, a comparison between the Fig.1-2 and Fig. 3-4 shows that the prediction of the THM model and the VM model differs only for $1-2\%$. 
A comparison between the Fig. 5 and Fig. 6 shows that the prediction, for what concerns the time evolution, of the THM model and the VM model differs $ \leq 10\%$ (for $L \neq 0$, $\Lambda=0$, $\Omega_0=0.2$).
I want to stress that even if in the parameter space considered, the THM model gives similar results to those of 
the continuous formation model, one cannot forget that it does not account properly for energy and mass accumulation during the early stages of cluster formation. In any case the central aim of this paper is to study the M-T relation and its evolution, the comparison between the THM and VM is only a secondary point.

\section*{Acknowledgments}
We are grateful to E. N. Ercan for stimulating discussions during
the period in which this work was performed and Y. Ek\c{c}si. A. Del Popolo would
like to thank Bo$\breve{g}azi$\c{c}i University Research
Foundation for the financial support through the project code
01B304.



\end{document}